\documentclass[a4paper]{jpconf}
\usepackage{graphicx}
\begin{document}
\title{Gamma/Hadron Separation in Imaging Air Cherenkov Telescopes Using Deep Learning Libraries TensorFlow and PyTorch}

\author{E B Postnikov$^{1,*}$, A P Kryukov$^1$,
	S P Polyakov$^1$, D A Shipilov$^2$, and D P Zhurov$^2$}
\address{$^1$ Lomonosov Moscow State University Skobeltsyn Institute of Nuclear Physics (MSU SINP), Leninskie gory 1(2), GSP-1, Moscow, 119991, Russia}
\address{$^2$ Irkutsk State University (ISU), Karl Marx street 1, Irkutsk, 664003, Russia}
\ead{$^*$evgeny.post@gmail.com\\
}

\begin{abstract}
In this work we compare two open source machine learning libraries, PyTorch and TensorFlow, as software platforms for rejecting hadron background events detected by imaging air Cherenkov telescopes (IACTs). Monte Carlo simulation for the TAIGA-IACT telescope is used to estimate background rejection quality. A wide variety of neural network algorithms provided by both libraries can easily be tested on various types of data, which is useful for various imaging air Cherenkov experiments. The work is a component of the Astroparticle.online project, which collaborates with the TAIGA and KASCADE experiments and welcomes any astroparticle experiment to join.
\end{abstract}

\section{Introduction}
\label{sec-1}
Gamma-rays have extremely low photon fluxes at very-high energies, 
so that the appropriate detector size would be too large for current space-based instruments, and, therefore, only ground-based measurements are accessible. With the imaging air Cherenkov technique \cite{1}, gamma-rays are observed on the ground optically via the Cherenkov light emitted by extensive showers of secondary particles in the air when a very-high-energy gamma-ray strikes the atmosphere. 

However, very energetic gamma-rays contribute only a minuscule fraction to the flux of cosmic rays (below one per million \cite{2}), and that is why a very important problem of data analysis is discrimination between gamma-rays and cosmic rays, which are hadrons (mostly protons), given a telescopic image of the particle shower.

The conventional way to solve this problem is a parametrization of the image using empirical variables named after Hillas \cite{3}: length, width, orientation, etc.
Their combinations, also empirical, serve as a discriminator between gamma-like and hadron-like images taking into account that on average gamma-rays produce narrower images than cosmic rays, and gamma-ray images are orientated towards the point of the gamma-ray source projection on an image plane whereas cosmic-ray images are randomly oriented.
For example, consecutive cuts on various Hillas parameters can be used to remove background events step-by-step, with optimal cut values determined by Monte Carlo simulation of the telescope \cite{4}.
Various attempts of combining these empirical parameters and derived quantities into machine learning algorithms have been carried out \cite{5, 6, 7, 8, 9}, but only three very recent efforts of a deep learning approach, which means fully automatic choice of the image features instead of empirical ones, have been made so far \cite{10, 11, 12}, of which the first one is an attempt not to distinguish primary gamma-rays from cosmic rays, but to select a special kind of images produced by secondary muons (`Muon Hunter' project \cite{13}). 

In this work we check the possibility of solving the above mentioned problem by Monte Carlo simulation of the TAIGA-IACT telescope \cite{14}.
The work is a component of the Karlsruhe-Russian astroparticle data life cycle initiative (also known as Astroparticle.online). This initiative aims to develop an open science system for collecting, storing, and analyzing astroparticle physics data. Currently it works with the TAIGA \cite{15} and KASCADE \cite{16} experiments and welcomes any other astroparticle experiments to join.


\section{Monte Carlo simulations}
\label{sec-2}
Simulation was performed for primary gamma-rays and protons of power-law energy spectra with the experimental value of the slope. The IACT pointing direction spanned the zenith angle range of 30--40$^\circ$ corresponding to the Crab Nebula observations of the TAIGA-IACT telescope. Protons, which are background cosmic ray particles in the imaging air Cherenkov technique, were incident within up to $\pm10^\circ$ around the fixed pointing direction of the IACT. Gamma-rays were incident within up to $\pm0.05^\circ$ around the IACT pointing direction, which is the expected IACT pointing and tracking accuracy.

At the first step, the shower development in the atmosphere was simulated with the CORSIKA package \cite{17}. The response of the IACT system was simulated at the second step using the OPTICA-TAIGA software developed for this task at JINR, Dubna \cite{18}. The mirror of the TAIGA-IACT consists of 29 segments and has an area of about 8.5 m$^2$ and a focal length of 4.75 m. The camera is located at the focus, it consists of 560 photomultipliers with the total field of view (FOV) diameter $\sim$9.6$^\circ$, and the single pixel FOV $0.36^\circ$. Cherenkov photons of the shower were traced through the IACT optical system, the number of corresponding photoelectrons (p.e.) in each pixel of the camera was counted, and the arrival time of each of them was calculated. 

Next, an output image was formed using a dedicated software developed at MSU SINP. For that purpose, photoelectrons in each photomultiplier were transformed into electric pulses using signal pulse-shape parametrization, and triggering and readout procedures were simulated for the data acquisition system. Upon readout, the Cherenkov signals are merged with night sky background (NSB) fluctuations, from a dedicated NSB simulation. A final image to be analyzed is obtained by subtracting the average NSB value (pedestal) from the pulse amplitudes (including NSB fluctuations) after readout in each photomultiplier.


\section{Image cleaning}
\label{sec-3}
The procedure of image reconstruction above NSB is called image cleaning \cite{19}. 
Conventional image cleaning procedure is two-parametric: it excludes from subsequent analysis all image pixels except the `core pixels', i.e. those with the amplitude above a `core threshold' and at least one neighbour pixel above a `neighbour threshold', and the neighbour pixels themselves.

\begin{figure}[h]
\begin{minipage}{20pc}
\includegraphics[width=18pc]{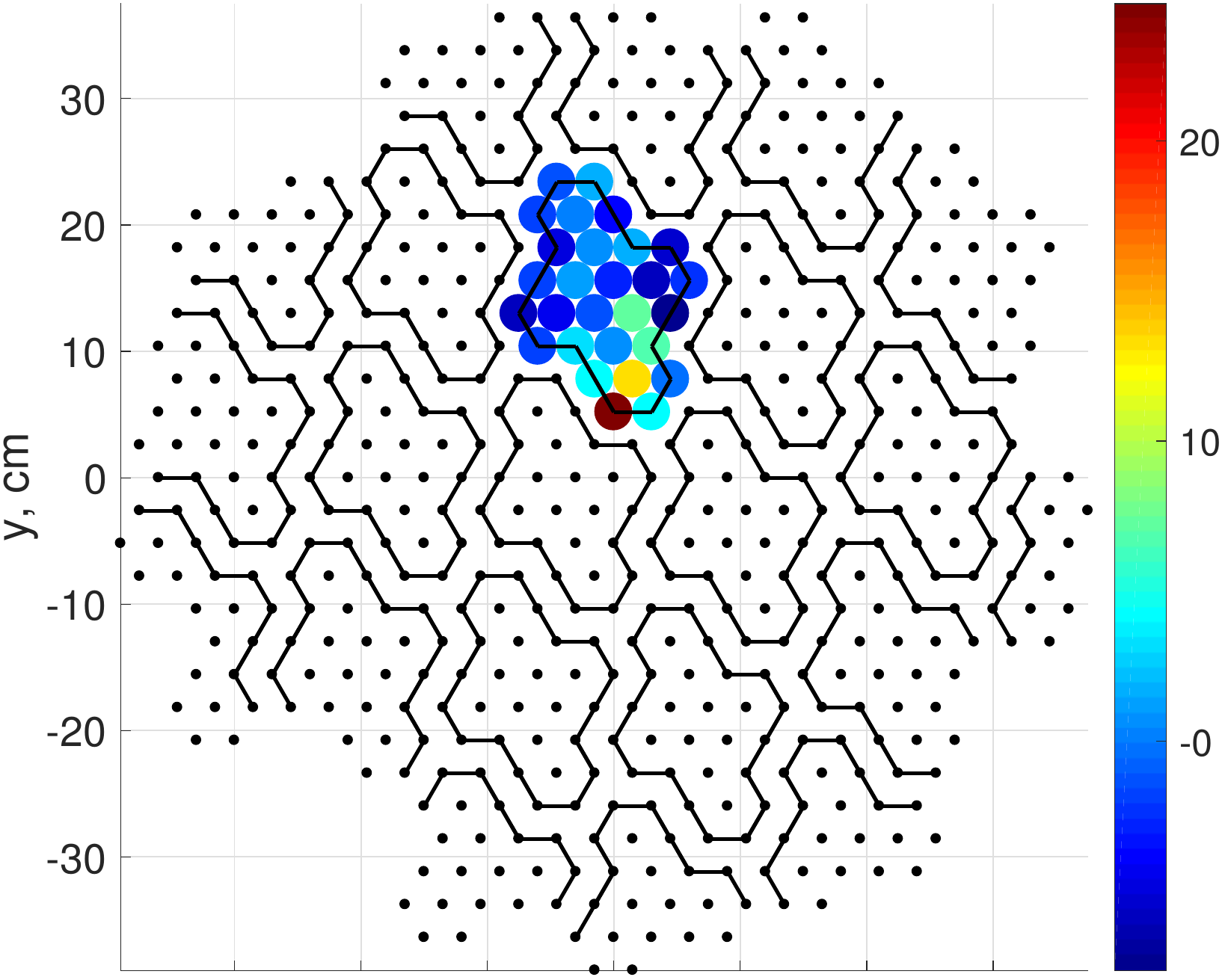}
\end{minipage}
\begin{minipage}{20pc}
\includegraphics[width=17.2pc]{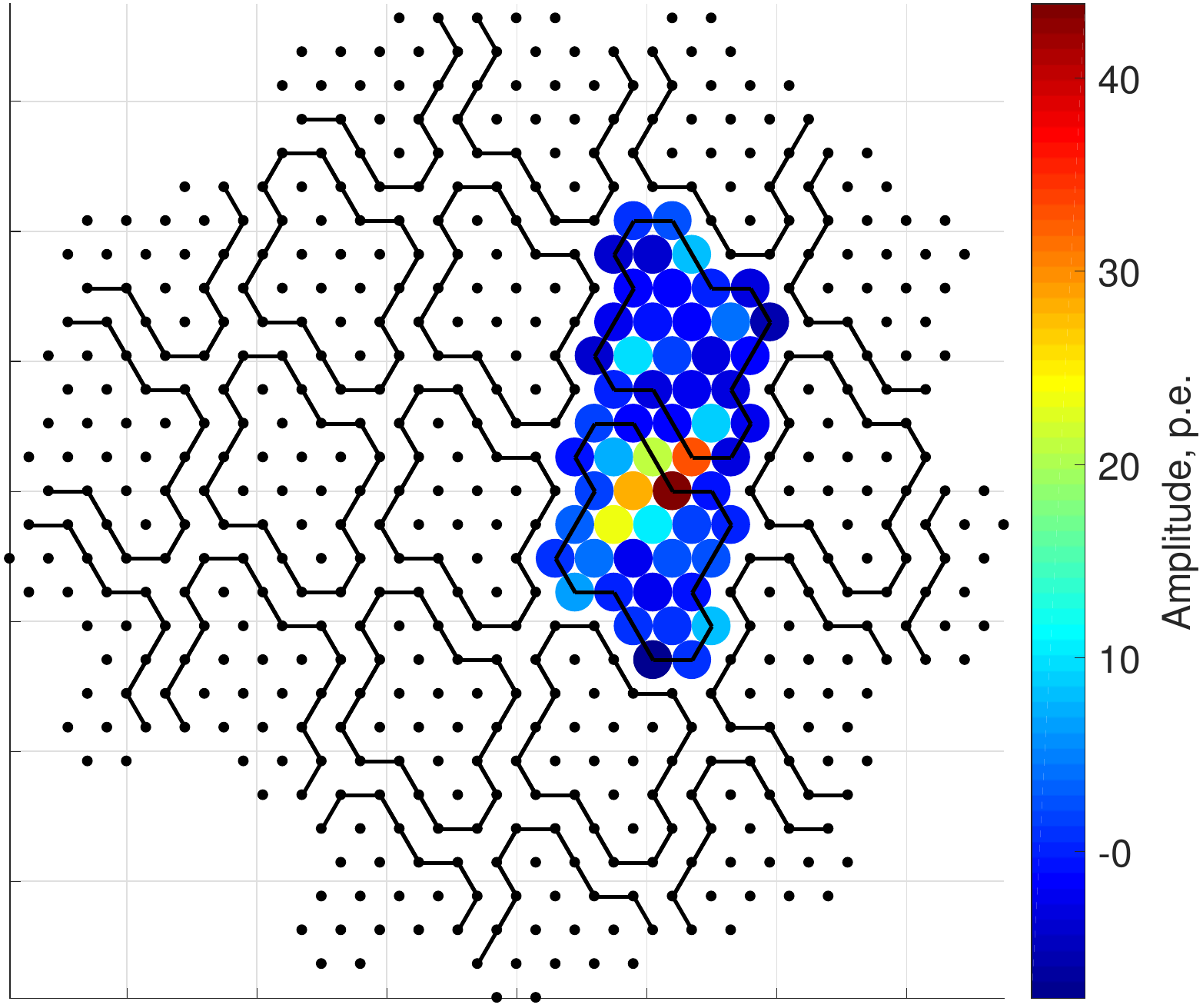}
\end{minipage} 

\begin{minipage}{20pc}
\includegraphics[width=18pc]{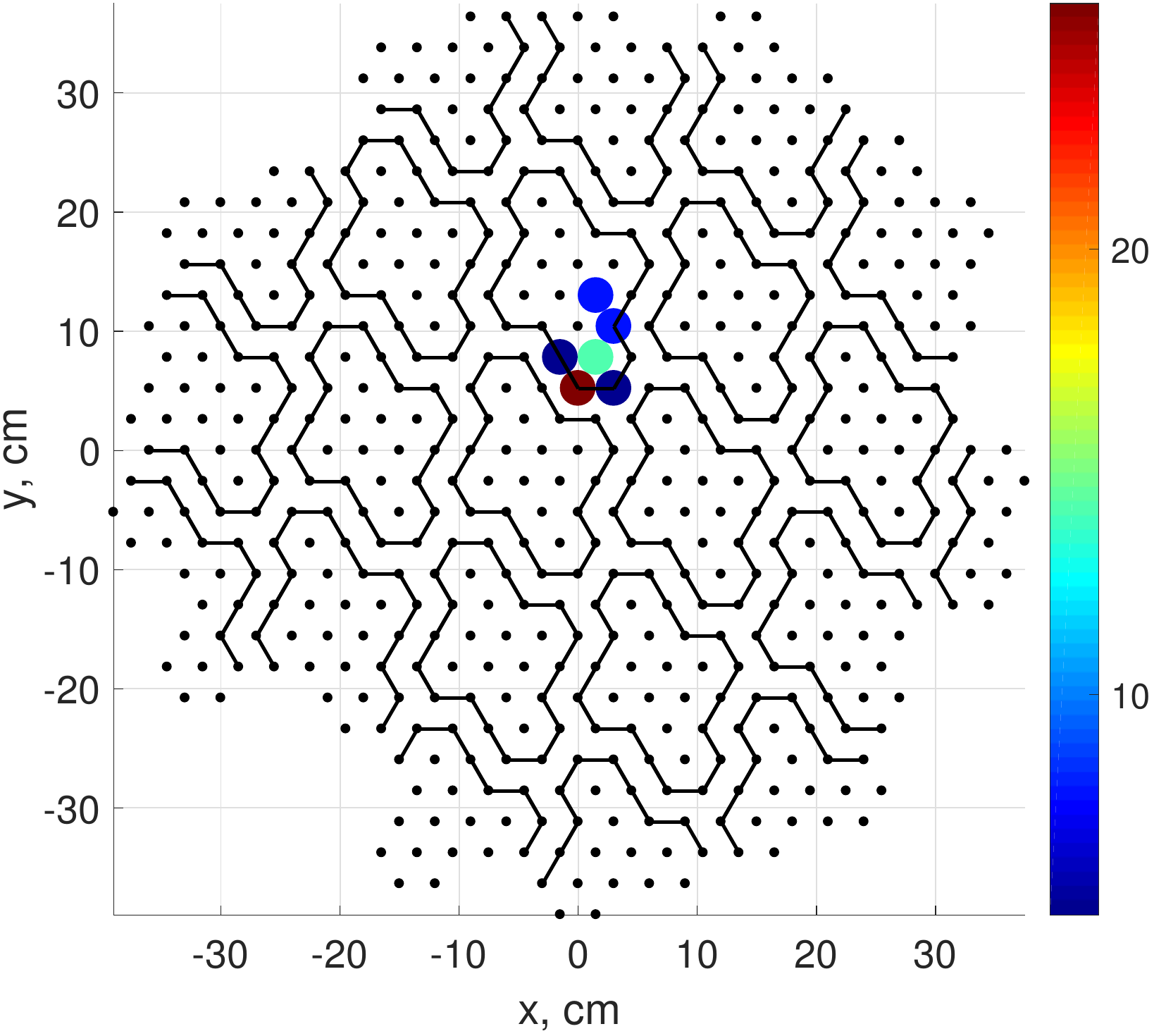}
\end{minipage}
\begin{minipage}{20pc}
\includegraphics[width=17.2pc]{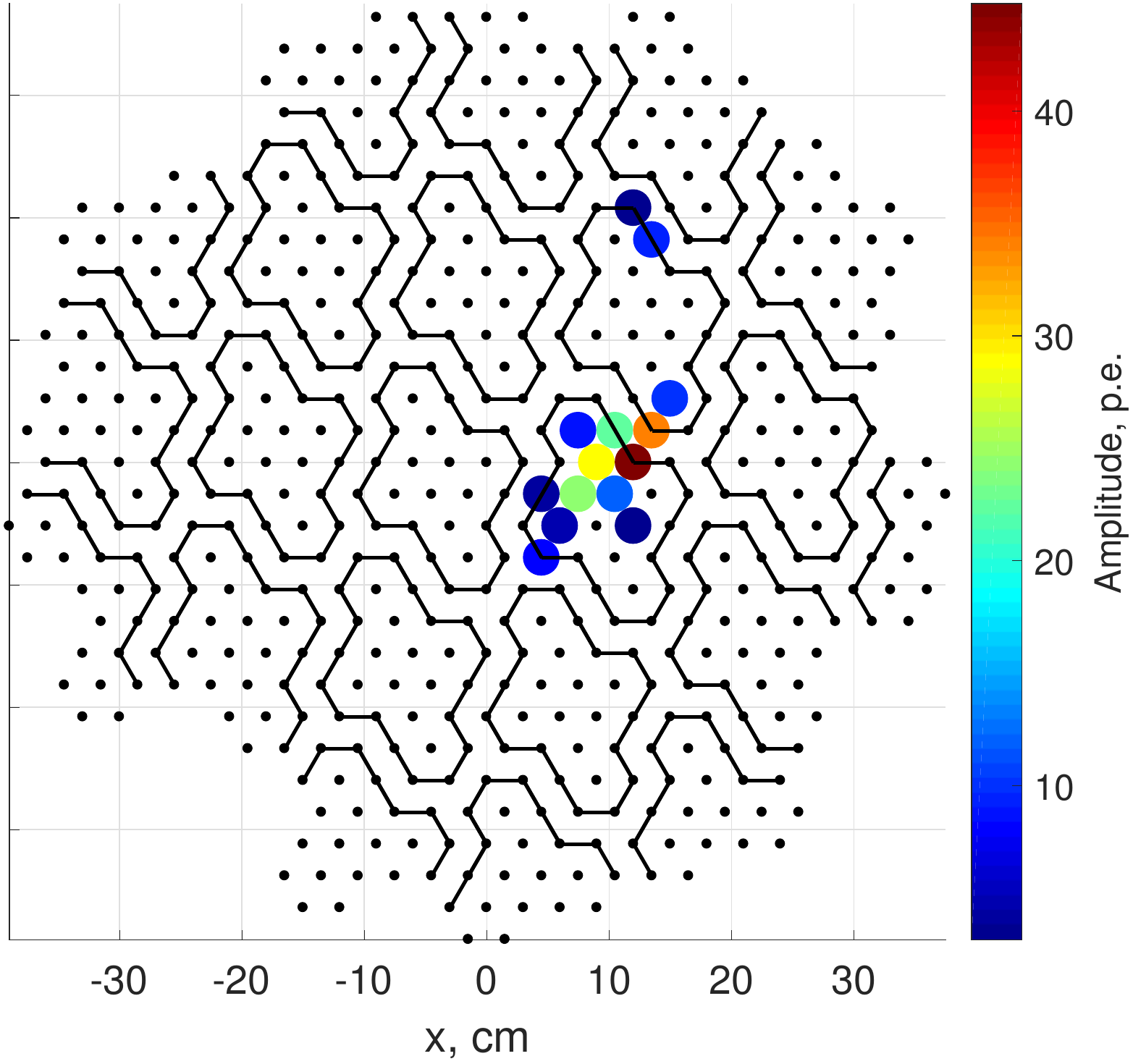}
\end{minipage} 
\caption{\label{figure1}Gamma-ray (left panel) and proton (right panel) images before image cleaning (top panel) after image cleaning with a low threshold (bottom panel).}
\end{figure}

An example of the simulated image before and after cleaning with a low threshold (core threshold is 6 p.e. and neighbour threshold 3 p.e., whereas an average NSB amplitude is 2.6 p.e.) is given in figure \ref{figure1}. Cluster structure of the telescope camera is also shown (black lines demarcate clusters of pixels), and the triggering procedure feature is also clearly visible: only the part of the image within the triggered clusters is registered. 

In this work we tried deep learning both with and without image cleaning. 

\section{Hillas parameters for comparison with the deep learning technique}
\label{sec-4}
For the reference technique, two Hillas parameters \cite{3} most informative for discrimination between gamma-rays and protons were calculated: the image width \textit{w} and angle $\alpha$ between the image orientation and the direction to the camera centre. Since on average gamma-rays produce narrower images than cosmic rays, and gamma-ray images are orientated towards the gamma-ray source projection on camera (in our case it's the centre of the camera, (0; 0) point in figure \ref{figure1}), these two parameters allow discriminating gamma-ray images (as those with small \textit{w} and $\alpha$) against proton background.

The reference gamma/proton separation technique to compare with the deep learning approach was a set of two consecutive cuts (on \textit{w} and $\alpha$) with the optimal cut values found as maximizing Q value (section \ref{sec-5.3}) on the learning samples of simulated gamma-rays and protons under the condition that the number of incorrectly identified gamma-rays is below 50\%. 




\section{Methods}
\label{sec-5}
\subsection{Convolutional neural network (CNN)}
\label{sec-5.1}
Under the deep learning approach, the so-called convolutional neural network (CNN) was chosen as it's a kind of artificial neural network that uses a special architecture, which is particularly well-adapted to classify images. Today, CNN or some close variant is used in most neural networks for image recognition, and it's also an independent choice for all other published deep learning applications to IACT technique \cite{10, 11, 12}. Besides its discriminative power, an advantage of CNN is a fully automatic algorithm, including automatic choice of image features (`capable of classifying IACT images without any prior parametrization', \cite{11}). The core principle behind CNN is that its convolutional layers apply a convolution operation (cross-correlation, or simply filtering) to the input, passing the result to the next layer, and so on. Special features of feedback (dropout \cite{20} etc.) avoid overfitting that was the problem for conventional artificial neural network.

CNN is realized in free convenient software packages: PyTorch \cite{21} and TensorFlow \cite{22}, and presently we are working with both of them. In contrast to work \cite{11}, pixels of the TAIGA-IACT camera are not square, but hexagonal. This special feature has not been fully taken into account yet, but only an approximation of the regular square grid using oblique coordinates with angle 60$^o$ was used.  

Training datasets contained gamma-ray and proton images (Monte Carlo of TAIGA-IACT, energy distributions in the range 2--60 and 3--100 TeV respectively with the spectral slope -2.6); NSB, trigger procedure and detector response added. 
Test datasets (different from training ones) of gamma-ray and proton images in random proportion (blind analysis) were classified by each of the packages: TensorFlow and PyTorch. The output of each of both packages is a scalar parameter, `probability' of being gamma or hadron; the term `hadronness' \cite{23} was introduced by the MAGIC collaboration: it `spans a range between 0 and 1 and characterizes the event images being less or more hadronlike').
\subsection{Quality criterion}
\label{sec-5.3}
As a quality criterion of particle separation the selection quality factor $Q$ was estimated. This factor indicates an improvement of a significance $S_0$ of the statistical hypothesis that the events do not belong to the background. For Poisson distribution (that is for a large number of events), the selection quality factor is:
\begin{equation}
Q=\epsilon_{\gamma}/\sqrt{\epsilon_{bckgr}},
\end{equation}
where $\epsilon_{\gamma}$ and $\epsilon_{bckgr}$ are relative numbers of gamma-ray events and background events after selection. For our task we consider protons as background and aim to select gamma-rays above this background. 
\subsection{Gamma/proton discrimination without image cleaning}
\label{sec-5.4}
This basic advantage of CNN made us first trying gamma-ray separation from proton background using Monte Carlo images without image cleaning (section \ref{sec-3}) at all. For that purpose training and test samples with neither cleaning nor preselection applied were given for analysis to both CNN packages (PyTorch, TensorFlow).

 A conventional Hillas analysis in a simplified version using only two basic cuts (section \ref{sec-4}): image width \textit{w} and angle $\alpha$, was used as a reference method. Conventional procedure of image cleaning (in our case, low-threshold cleaning with 6 p.e./3 p.e. threshold values) was also applied to the data before applying cuts; however, the images that were completely removed by image cleaning were also taken into account as `protons' -- for a fair comparison.
\subsection{Gamma/proton discrimination after image cleaning}
\label{sec-5.5}
Training and test samples after image cleaning with low thresholds (6 p.e./3 p.e.), were also analyzed by CNN in both software packages. In case the image had less than 3 pixels after cleaning, it was removed from the analysis before calculating quality factor. The same reference method as in the previous section was used with a slight difference that now the images removed by image cleaning were not taken into account -- as well for a fair comparison of both techniques.
\section{Results}
\label{sec-6}
Quality factor (1) values obtained by the best CNN configuration among all the trained networks are assembled in table \ref{ex} together with the quality factor for the reference technique (sections \ref{sec-5.4}, \ref{sec-5.5}). Quality of the reference method of separation changed only slightly, because the reference technique always includes image cleaning, so the difference was only in the test sample size -- the first table row contains the value for all events, whereas the last row contains the value for all events except those removed after image cleaning (section \ref{sec-3}).
\begin{table}[h]
\caption{\label{ex}Quality factor for gamma/proton separation.}
\begin{center}
\begin{tabular}{llll}
\br
&Reference(Hillas analysis)&CNN(PyTorch)&CNN(TensorFlow)\\
\mr
Without image cleaning&1.76&1.74&1.48\\
With image cleaning&1.70&2.55&2.99\\
\br
\end{tabular}
\end{center}
\end{table}
\section{Discussion}
A fully automatic algorithm without image cleaning demonstrated the same quality as the simplest version of the conventional technique, Hillas analysis \cite{3}. This means that one of more elaborated versions of this technique \cite{5, 6, 7, 8, 9} would definitely overwhelm our CNN result. Therefore, some image cleaning procedure is necessary before the CNN implementation; however, the one we used is not optimal for CNN quality. Besides, some more elaborated CNN can also be applied to this task in future, so the conclusion of image cleaning necessity before CNN application should be considered preliminary.

\begin{figure}[h]
\includegraphics[width=37.5pc]{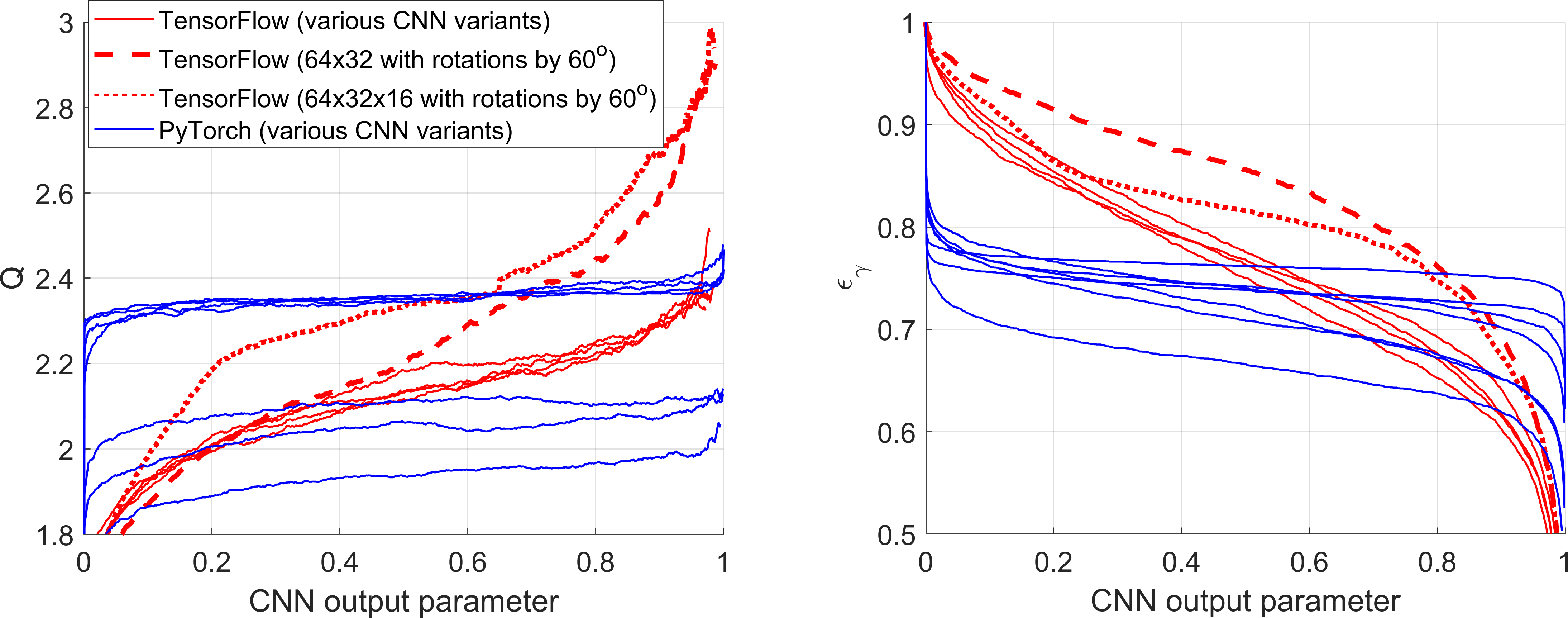}
\caption{\label{figure2}Quality factor (left) and percentage of correctly identified gamma-rays (right) vs CNN output parameter.}
\end{figure}

Comparison of different CNN versions for both CNN packages (PyTorch and TensorFlow) is illustrated in figure \ref{figure2}. Overall PyTorch had more stable results in a wide range of CNN output parameter values (section \ref{sec-5.1}). However, the TensorFlow CNN version trained by a `modified' training sample, which contains additional images obtained by rotating 70\% of images from the initial sample by 60$^o$ (symmetry angle of hexagonal structure), and thereby has $\sim$180000 events instead of $\sim$30000, shows significant improvement. Therefore, we think that these packages demonstrate approximately equal performance, and training sample size has the strongest influence on it. (PyTorch CNN has not been tested with the modified training sample yet.) Another issue is a problem of choosing the output parameter value; to get better quality, this value should be as high (close to 1) as possible (figure \ref{figure2} (left)), but not as high as to lead to misidentifying more than 50\% of gamma-rays (that happens when the curves in figure \ref{figure2} (right) abruptly fall down below the $\epsilon_{\gamma}$=0.5 value).
\label{sec-7}

\section{Conclusions}
\label{sec-8}
Both CNN packages were tested and showed approximately equal results. Two algorithms were found to significantly improve CNN separation quality: image cleaning, even with a low threshold (section \ref{sec-3}), and additional image rotation in training sample, which allows increasing sample size. Problem of choosing the CNN output parameter value is still unsolved; however, there is still considerable potential to further improve the results by taking into account the hexagonal pixel shape and increasing training sample size by one order of magnitude.  
\ack
The work was supported by the Russian Science Foundation, grant \#18-41-06003.

\end{document}